\documentclass[%
reprint,
superscriptaddress,
 amsmath,amssymb,
 aps,
]{revtex4-1}

\usepackage{graphicx,setspace}
\usepackage{dcolumn}
\usepackage{bm}
\usepackage[mathlines]{lineno}
\usepackage[colorlinks=true,citecolor=blue,linkcolor=blue,anchorcolor=green, anchorcolor=green,urlcolor=blue]{hyperref}
\usepackage{mathrsfs}
\usepackage{xcolor}
\usepackage{float}
\usepackage{ulem}
\usepackage[utf8]{inputenc}

\newcommand{\be}{\begin{equation}}
\newcommand{\ee}{\end{equation}}
\newcommand{\bal}{\begin{aligned}}
\newcommand{\eal}{\end{aligned}}

\newcommand{\bea}{\setlength\arraycolsep{2pt} \begin{eqnarray}}
\newcommand{\eea}{\end{eqnarray}}

\graphicspath{{Bilder/}}

\begin{document}


\title{Revisiting collisional Penrose processes in term of escape probabilities for spinning particles}

\author{Ming Zhang}
\email{mingzhang@jxnu.edu.cn}
\affiliation{Department of Physics, Jiangxi Normal University, Nanchang 330022, China}
\author{Jie Jiang}
\email{jiejiang@mail.bnu.edu.cn (corresponding author)}
\affiliation{Department of Physics, Beijing Normal University, Beijing 100875, China}


\begin{abstract}
We first study the escape probability of the spinning particle emitted from the Kerr black hole and find that the escape probability increases with the spin of the particle around the extreme Kerr black hole; in contrast, the escape probability decreases at the position near the horizon but increases at the position far away from the horizon with the increasing spin of the particle. We then probe the relation between the escape probabilities and the energy extraction efficiencies of collisional Penrose processes for the particles with varying spin. For the extreme Kerr black hole, the efficiency increases with the escape probability; for the non-extreme Kerr black hole, the near-horizon-efficiency decreases with the escape probability whilst the efficiency may increase with the escape probability in the ergosphere. In the event horizon limit, we also find that the average escape probability of the spinning particle produced in the collisional Penrose process decreases with the rotation parameter of the Kerr black hole.
\end{abstract}


\maketitle


\section{Introduction}
The well-known Penrose process (PP) \cite{Penrose2002} states that an escaped particle can carry more energy than the one from whom it is disintegrated from in a background of Kerr spacetime. A more realistic scenery overcoming the seeming implausibility of disintegration in PP is believed to be the collisional Penrose process (CPP1) \cite{piran1975high}, where two particles plunging into the ergoregion and collide. The energy extraction efficiency in CPP, however, was verified to be as qualitatively similar to PP \cite{Piran1977dm}. There is another process (CPP2), where one of the infalling particles with sufficient angular momentum turns around the rotating black hole and collides on its outgoing orbit with the other infalling particle \cite{Schnittman:2014zsa}, can work with relatively higher efficiency. Furthermore, the super Penrose process (SPP) \cite{Berti:2014lva,Zaslavskii:2015fqy,Zaslavskii:2019pdc}, where a head-on collision takes place between one outgoing particle and one ingoing particle, can reach an infinite efficiency at the horizon limit.

The Banados-Silk-West (BSW) mechanism, which states that the centre-of-mass energy for two spinning particles (one with a critical angular momentum) can be arbitrarily high after a collision near the horizon of an extreme rotating black hole \cite{Banados:2009pr,Armaza:2015eha,Guo:2016vbt}, have been renovating the investigation of energy extraction from the black hole, as the effect of spin carried by the collisional particles was discussed qualitatively \cite{Liu:2018myg,Mukherjee:2018kju} and quantitatively \cite{Maeda:2018hfi} while the effect of the charge carried by the collisional particles was introduced to some extent \cite{Zhang:2018gpn}.

In fact, the observability of the collisional events around the black hole depends on how often a particle can escape from the black hole to spatial infinity \cite{Ogasawara:2019mir}. The astrophysical process in the strong gravity field of the  black hole can be further understood by using the notion of the escape probability for the particle, by which we can know which portion of the radiation emitted from the particle source is trapped whilst the complementary portion can escape to spatial infinity \cite{Schee:2008kz}.  In this paper, we will first briefly review the equations of motion for the spinning particle in the Kerr spacetime in Sec. \ref{eos} for later requirement. Then we will calculate the escape probability for the spinning particle which is supposed to emit isotropically from a particle source in Sec. \ref{epxx}. In Sec. \ref{epcpp}, we will investigate the relation between the energy extraction efficiency of the collisional Penrose process and the escape probability of the produced particle. Sec. \ref{con} will be devoted to our conclusions.

\section{The equations of motion for a spinning particle in the Kerr spacetime}\label{eos}
The motion of an astronomical test particle whose pole/de-pole moment is considered in curved spacetime can be described by the well-known Mathisson-Papapetrou-Dixon (MPD) equations \cite{Hojman:1976kn}
\begin{equation}
\frac{D P^a}{D \tau }=-\frac{1}{2}R^a{}_{bcd}v^{b} S^{cd},
\end{equation}
\begin{equation}
\frac{D S^{\text{ab}}}{D \tau }=2P^{[a}v^{b]},
\end{equation}
where $\tau$ is the parameter along the world line of the particle. $R_{abcd}$ is the Riemannian curvature tensor of the spacetime geometry. The four-momentum $P^{a}$ is related to the particle's mass $\mathcal{M}$ by \cite{Costa:2017kdr,Obukhov:2010kn}
\begin{equation}
 P^a  P_a=-\mathcal{M}^{2}
 \end{equation}
 in the zero three-momentum frame and it together with the particle's four-velocity $v^{a}$ also defines the other mass $m$ in the zero three-velocity frame by \cite{Costa:2017kdr,Obukhov:2010kn}
 \begin{equation}
P_av^a=-m.
 \end{equation}
 The normalized four-momentum of the particle is
 \begin{equation}
u^a\equiv \frac{P^a}{\mathcal{M}}.
\end{equation}
The dynamical mass of the particle can be ensured to be conserved by using the well-known Tulczyjew condition \cite{tulczyjew1959motion,dixon1964covariant,Dixon:1970zza}
\begin{equation}
S^{ab} P_b=0.
\end{equation}
Also, the magnitude of the spin $S$ can be invariable in condition of  \cite{Wald:1972sz}
\begin{equation}
S^{ab}S_{ab}=2 S^2.
\end{equation}
As $\mathcal{M}=m+\mathcal{O}(S^{2})$  \cite{Ruangsri:2015cvg}, we can have $v^{a}u_{a}=-1$ by reparameterizing $\tau$ \cite{Tanaka:1996ht,Saijo:1998mn,Mukherjee:2018zug} . Accordingly, the four-momentum of particle can be obtained as  \cite{Hojman:1976kn,Costa:2017kdr,Obukhov:2010kn,Lukes-Gerakopoulos:2017cru}
\begin{equation}\label{relationuv}
v^a=u^a+\frac{2 S^{ab} u^c R_{bcde} S^{de}}{S^{bc} R_{bcde} S^{de}+4 \mathcal{M}^2}.
\end{equation}

We consider the Kerr spacetime in this paper. After choosing the unit $c=G=1$, the Kerr line element can be written in Boyer-Lindquist coordinates as
\begin{equation}\label{metric}
\begin{aligned}
ds^2=&-\left(1-\frac{2 Mr}{\Sigma }\right)dt^2 +\frac{\Sigma }{\Delta}dr^2+ \Sigma d\theta^2\\&-\frac{4M a r}{\Sigma }\sin ^2\theta dt d\phi+\frac{\Xi }{\Sigma }\sin ^2\theta d\phi^2,
\end{aligned}
\end{equation}
where
\begin{align}
\Sigma &=r^2+a^2 \cos ^2\theta,\nonumber~\\ \Delta &=r^{2}-2 Mr+a^2\nonumber,\\
\Xi &=\left(a^2+r^2\right)^2-a^2 \Delta  \sin ^2 \theta\nonumber.
\end{align}
$M$ is the mass of the black hole and $a$ is the rotation parameter defined by $J/M$ with $J$ the angular momentum of the black hole. The event horizon and the  stationary limit  locate at
\begin{equation}
r_+=M+\sqrt{M^2-a^2},
\end{equation}
\begin{equation}
r_e=M+\sqrt{M^2-a^2\cos\theta^{2}}.
\end{equation}
The collisional Penrose processes that will be discussed in Sec. \ref{epcpp} take place in the ergosphere $r_{+}<r_{*}<r_{e}$. The Kerr spacetime admits conserved energy and angular momentum for the particle as
\begin{equation}
e=\frac{1}{2 \mathcal{M}}S^{tb} \nabla _b\xi _t-\xi _t u^t,
\end{equation}
\begin{equation}\label{conj}
j=-\frac{1}{2 \mathcal{M}}S^{\phi b} \nabla _b\xi_\phi+u^\phi \xi_\phi,
\end{equation}
where
$$
\xi_{t}\equiv \left(\frac{\partial }{\partial t}\right)^a,\quad\xi_{\phi}\equiv \left(\frac{\partial }{\partial \phi }\right)^a.
$$
In an orthogonal normalized tetrad $\{e_{a}^{(\mu)}\}$, which reexpresses  the metric (\ref{metric}) as
\begin{equation}
ds^2=\eta_{(i)(j)}e_{a}^{(i)}e_{b}^{(j)},
\end{equation}
with $\eta_{(i)(j)}=\text{diag}(-1,1,1,1)$, we can introduce the spin vector $s^{(a)}$ of the particle as
 \begin{equation}
S^{(c) (d)}=\mathcal{M}\varepsilon^{(c)(d)}_{}{}_{(a) (b)}u^{(a)}s^{(b)},
\end{equation}
where we have the completely antisymmetric tensor ${\bm{\varepsilon}}$ as $\varepsilon_{(0)(1)(2)(3)}=1$. Considering that the motion of the spinning particle is confined in the equatorial plane, we then only have a nonvanishing component of the spin vector $s^{(2)}=-s$, with $s$ the magnitude of the spin and $s>0$ corresponds to a spin direction parallel to that of the Kerr black hole. As a result, we get nonvanishing components of the spin tensor $S^{(a)(b)}$ as
\begin{subequations}
\begin{align}
S^{(0)(1)}=&-\mathcal{M} s u^{(3)},\\
S^{(0)(3)}=&\mathcal{M} s u^{(1)},\\
S^{(1)(3)}=&\mathcal{M} s u^{(0)}.
\end{align}
\end{subequations}

\begin{figure*}[!htbp] 
   \centering
   \includegraphics[width=2.3in]{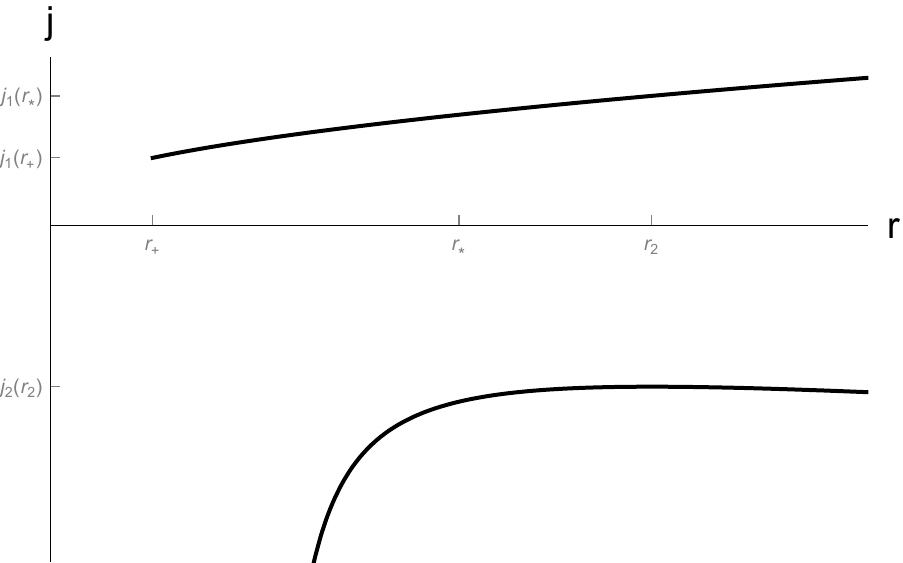}
    \includegraphics[width=2.3in]{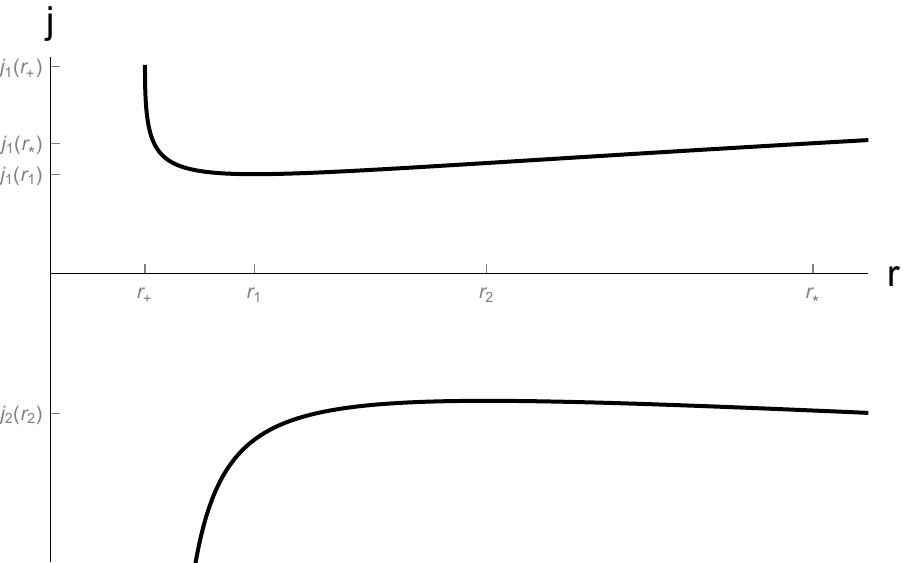}
     \includegraphics[width=2.3in]{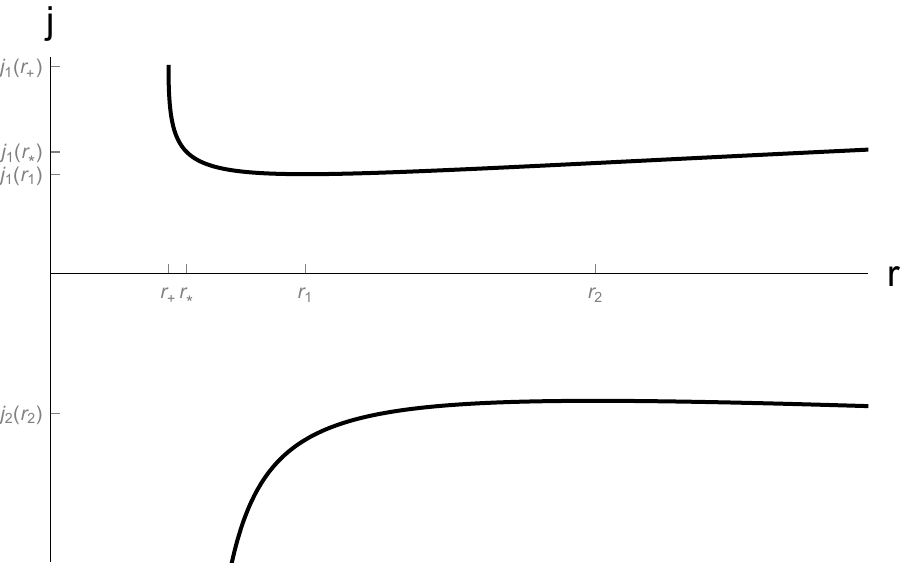}
   \caption{The critical angular momentum which makes $O=0$. The left diagram corresponds to the extreme Kerr black hole case. The middle and right diagrams are for the non-extreme Kerr black hole cases where $r_*>r_1$ and $r_*<r_1$, respectively.}
   \label{schematic}
\end{figure*}

We now choose the Carter frame to calculate the equations of motion for the spinning particle in the Kerr spacetime. In the Carter frame
\begin{subequations}
\begin{align}
e_a^{(0)}&=\sqrt{\frac{\Delta }{\Sigma }} \left(dt-a \sin^{2}\theta d\phi \right),\\e_a^{(1)}&=\sqrt{\frac{\Sigma }{\Delta }}dr,\\e_a^{(2)}&= \sqrt{\Sigma }d\theta,\\e_a^{(3)}&=\frac{\sin \theta}{\sqrt{\Sigma }} \left[-a dt+\left(a^2+r^2\right)d\phi\right],
\end{align}
\end{subequations}
the normalized four-momentum of the spinning particle is 
\begin{equation}\begin{aligned}
u^{(0)}=\frac{\left[e r^5+(ea+es-j)a r^3+  (a e M-j M)sr^{2}\right]}{\sqrt{\Delta}\mathcal{X}},\\
\end{aligned}\end{equation}
\begin{equation}
u^{(3)}=\frac{r^3 (j-ea-es)}{\mathcal{X}},
\end{equation}
\begin{equation}\label{ramo}
u^{(1)}=\sigma \sqrt{-1+(u^{(0)})^2  - (u^{(3)})^2}=\sigma \sqrt{O},
\end{equation}
where $\mathcal{X}=r^{4}-M r s^2$, $\sigma=1$ corresponds to a radially outgoing particle and $\sigma=-1$ for a radially ingoing one, $O$ is the radial effective potential of the particle. By using (\ref{relationuv}), we can obtain the 4-velocity of the spinning particle as 
\begin{align}
v^{(0)}&=\frac{ r^{4}-M s^2 r}{-3M r\left(u^{(3)}\right)^{2} -s^2 M r+r^4}u^{(0)},\\
v^{(1)}&=\frac{r^{4}-M s^2 r}{-3M r\left(u^{(3)}\right)^{2} -s^2 M r+r^4}u^{(1)},\label{proone}\\
v^{(3)}&=\frac{r^4+2 M r s^2}{-3M r\left(u^{(3)}\right)^{2} -s^2 M r+r^4}u^{(3)}.
\end{align}
The equations of motion for the spinning particle is \cite{Saijo:1998mn}
\begin{equation}\label{eos1}
\frac{dt}{d\tau}=\frac{\mathcal{X} \left(a^2 \mathcal{P}_2 \mathcal{X}+a \Delta  r\mathcal{P}_3 +\mathcal{P}_2   r^2 \mathcal{X} \right)}{\sqrt{\Delta}  \left[ -3 M {\mathcal{P}_1}^2  s^2 r^{5}+ {\mathcal{X}}^2 r^{4}-M r \mathcal{X}^2 s^2 \right]},
\end{equation}
\begin{equation}\label{rtau}
\frac{dr}{d\tau}=\sqrt{\frac{\Delta}{\Sigma}}v^{(1)},
\end{equation}
\begin{equation}\label{phitau}
\frac{d\phi}{d\tau}=\frac{1}{a \sin\theta^2}\left(\frac{dt}{d\tau}-\sqrt{\frac{\Sigma}{\Delta}}v^{(0)}\right),
\end{equation}
where
\begin{align}
\mathcal{P}_1&=r \left[j-e (a+s)\right]\nonumber,\\
\mathcal{P}_2 &=a^2 e r^2-a \left[e s \left(-Mr+r^{2}\right)+j r^{2}\right]\nonumber+r^4 e -j s M r\nonumber,\\
\mathcal{P}_3&=  2 M r s^2+r^4.\nonumber
\end{align}
When $s=0$, they reduce to the equations of motion for a spinless massive particle.

To make the motion of the spinning particle physical, we should constrain the ranges of the parameters. The particle's motion should comply with the time-like condition $v_{(a)}v^{(a)}<0$ and the forward-in time condition $dt/d\tau > 0$ \cite{Grib:2013hxa}. Besides, we should keep $s\lesssim r_0\ll r_+ \leqslant 2M, $ \cite{Wald:1972sz}, where $r_0$ denotes the size of the particle. Based on the necessity of the physical reasonability, we also restrict the radial effective potential $O\geqslant 0$. Starting from this condition, it can be proved that only the particle with a conserved angular momentum $j\leqslant 2e\equiv j_c$ can reach the horizon in the extreme Kerr geometry. A particle is critical if it holds an angular momentum $j=j_c$. Otherwise, the cases $j<j_c$ and $j>j_c$ correspond to a sub-critical particle and a super-critical particle, respectively. We will exhaustively discuss the non-extreme case in what follows.

\begin{figure*}[!htbp] 
   \centering
       \includegraphics[width=2.1in]{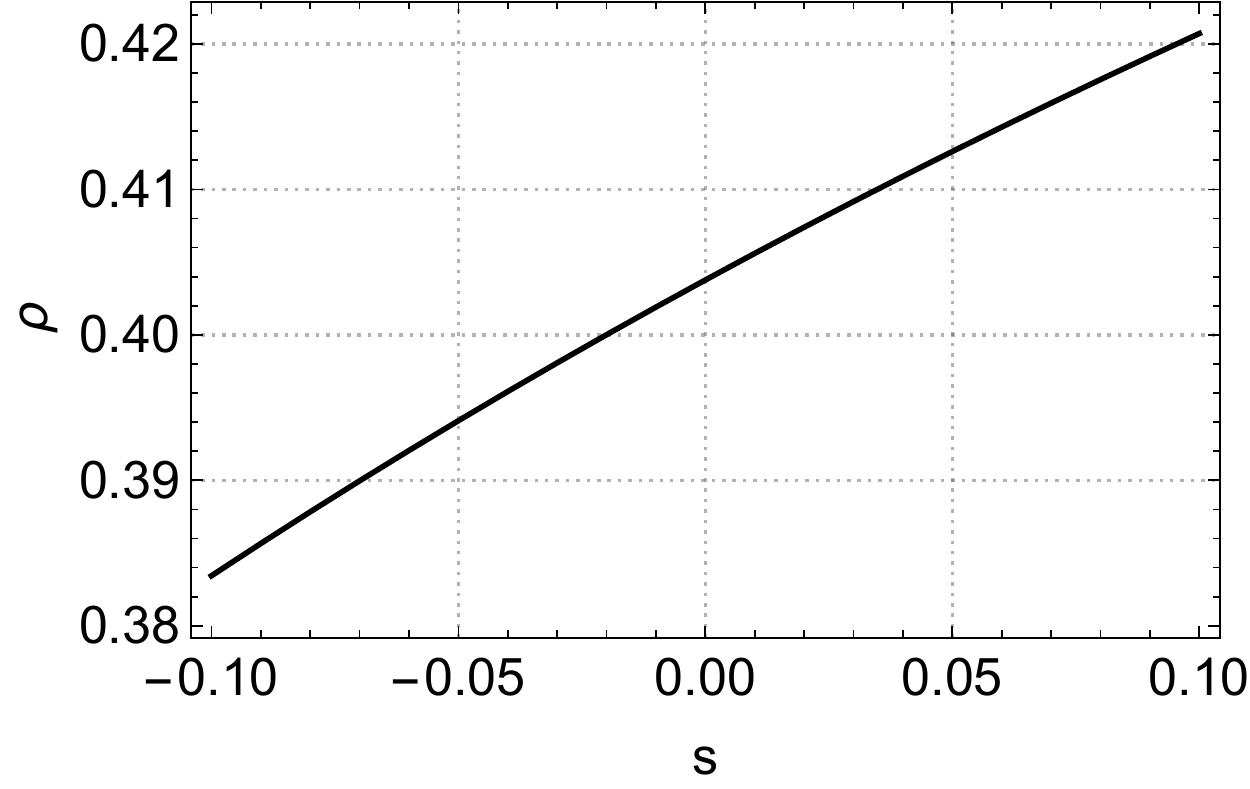}
   \includegraphics[width=2.1in]{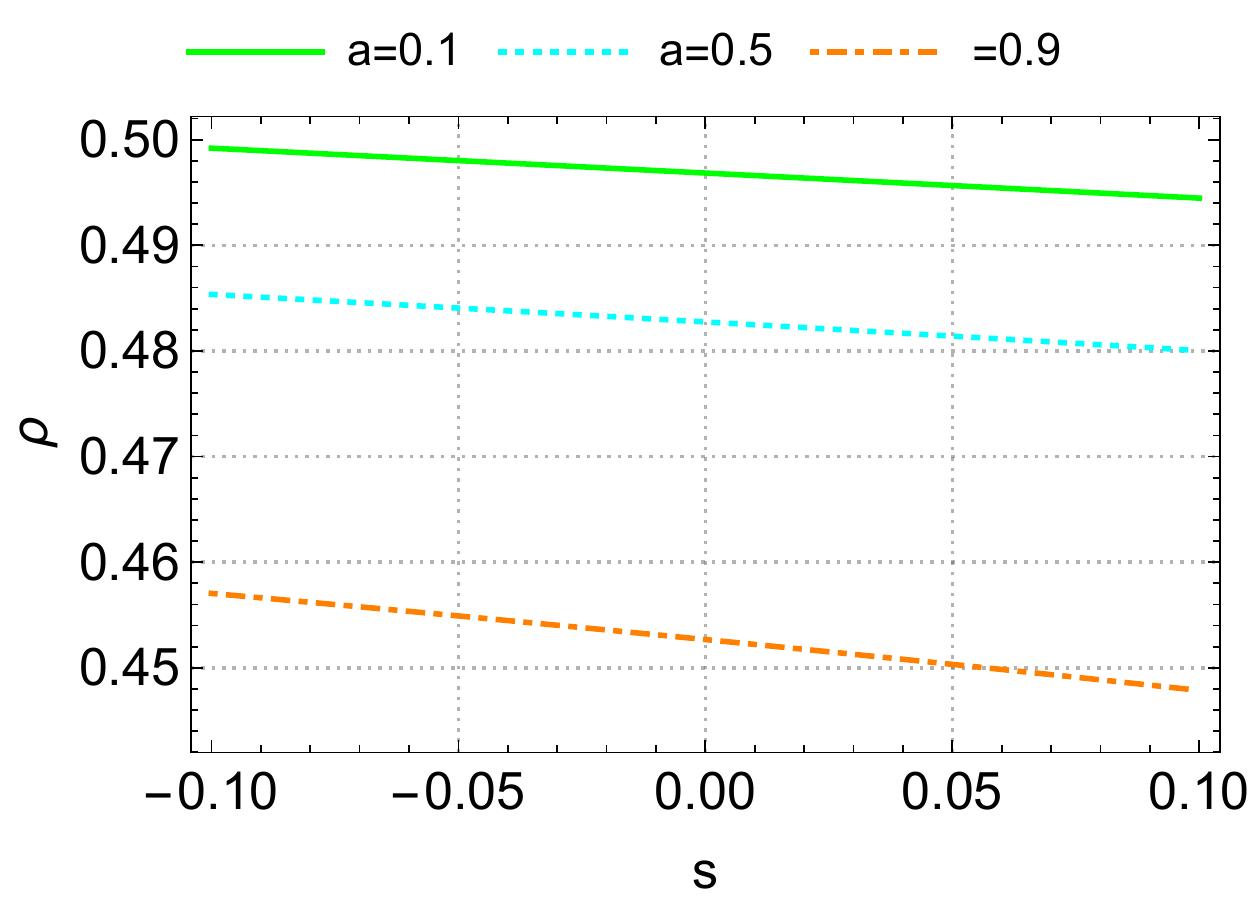}
      \includegraphics[width=2.1in]{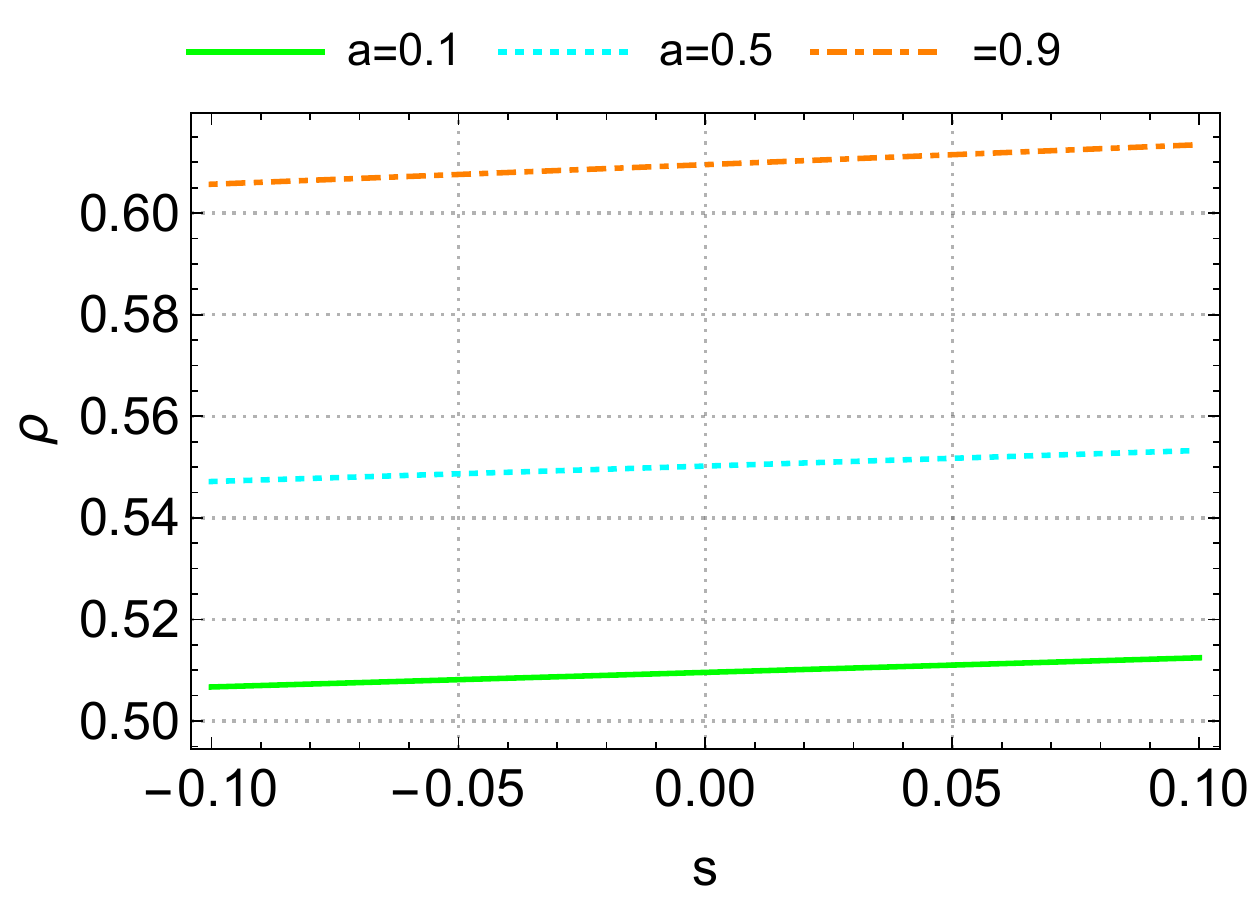}
   \caption{Escape probabilities of spinning particles from the Kerr black hole for $M=1,\,e=1,\, $$r_{*}=1.01r_{+}<r_{1}$ (left and middle diagrams), $r_{*}=10 r_{+}>r_{1}$ (right diagram). The left one is for the extreme Kerr black hole case and the others are for the non-extreme Kerr black hole case.}
   \label{ep}
\end{figure*}

\section{The escape probability of the spinning particle}\label{epxx}
Not loss of generality, we now set the Kerr black hole mass as $M=1$. As the radial and angular equations of motion for the spinning particle can not be separated, we here consider the escape probability of the spinning particle in the equatorial plane \cite{Mukherjee:2019jhd,carter1968global}. By solving $O=0$, we can obtain the critical conserved angular momenta of the particles as
\begin{equation}
j_{+}=\frac{a^2 \left(\mathcal{Y}_1-2 e r^2 s\right)+e (r-3) r^4 s+(r-2) r \mathcal{Y}_1-\mathcal{Y}_2}{r \left(-2 a r s+(r-2) r^3-s^2\right)},
\end{equation}
\begin{equation}
j_{-}=\frac{-a^2 \left(2 e r^2 s+\mathcal{Y}_1\right)+e (r-3) r^4 s-(r-2) r \mathcal{Y}_1-\mathcal{Y}_2}{r \left(-2 a r s+(r-2) r^3-s^2\right)},
\end{equation}
where
\begin{equation}
\mathcal{Y}_1= \left(r^3-s^2\right) \sqrt{\frac{2 a r s+2 r^3+s^2}{a^2+(r-2) r}},
\end{equation}
\begin{equation}
\mathcal{Y}_2=a e r \left(2 r^3+(r+1) s^2\right).
\end{equation}
At the event horizon limit, the two branches  join to one point
\begin{equation}
\begin{aligned}
j_{+}(r=r_+)&=j_-(r=r_+)\\&=\frac{e\left(a^4 s+2 a^3 r_+-a^2 r_+ s+a r_+ s^2+4 r_+ s\right)}{a^4+2 a s+s^2}.
\end{aligned}
\end{equation}
As shown in Fig. \ref{schematic}, we denote the minimal value of the critical angular momentum $j_{+}$ for the spinning particle as $j_+(r_{1})$ with $r_{1}$ the corresponding radial position, we also denote the maximal value of the critical angular momentum  $j_{-}$ as $j_-(r_{2})$ with $r_{2}$ the corresponding radial position. We set the particle source at the position $r_{*}$. If the Kerr black hole is extreme, we can have $r_{+}<r_{*}$; if the Kerr black hole is non-extreme, we have $r_{+}<r_{1}<r_{*}$ or $r_{+}<r_{*}\leqslant r_{1}$. In the background of the extreme Kerr black hole, the spinning particle at the position $r_{*}$ can escape to spatial infinity irrespective of its initial velocity if $j_{+}(r_+)<j<j_+(r_{*})$, and it can escape to spatial infinity only with initially outgoing velocity in condition of $j_-(r_{2})<j<j_{+}(r_+)$.  In the background of the non-extreme Kerr black hole, if the spinning particle with $j_+(r_{1})<j<j_+(r_{*})$ is located at $r_{*}>r_{1}$, it can escape to spatial infinity irrespective of the sign of its initial velocity,  and the outgoing particle with $j_-(r_{2})<j<j_+(r_{1})$ can emit to spatial infinity. In condition of $r_{*}\leqslant r_{1}$, no particle with initial ingoing velocity can go to spatial infinity and only outgoing particle with $j_-(r_2)<j<j_+(r_1)$ can escape to spatial infinity.

%

The emission angle $\alpha$ can be introduced for the particle produced by a source at the Carter's frame, which can be defined by the spinning particle's four-momentum as \cite{Schee:2008kz,Stuchlik:2018qyz}
\begin{equation}
\begin{aligned}
\sin\alpha=\frac{p^{(\phi)}}{\sqrt{\left(p^{(r)}\right)^2+\left(p^{(\phi)}\right)^2}},
\end{aligned}
\end{equation}
\begin{equation}
\begin{aligned}
\cos\alpha=\frac{p^{(r)}}{\sqrt{\left(p^{(r)}\right)^2+\left(p^{(\phi)}\right)^2}}.
\end{aligned}
\end{equation}
The critical angles that the spinning particle can escape to spatial infinity are
\begin{align}
\alpha_{\rm I}
&\equiv\alpha[\sigma=-1,\;j=j_{+}(r_+)],~~\text{extreme case}\\
\alpha_{\rm I}
&\equiv\alpha[\sigma=-1,\;j=j_+(r_{1})],~~\text{non-extreme case}\\
\alpha_{\rm II}
&\equiv\alpha[\sigma=-1,\;j=j_+(r_*)],\\
\alpha_{\rm I\hspace{-.1em}I\hspace{-.1em}I}
&\equiv\alpha[\sigma=1,\;j=j_+(r_*)],\\
\alpha_{\rm IV}
&\equiv\alpha[\sigma=1,\;j=j_{-}(r_2)].
\end{align}
Note that in case of $r_{*}<r_{1}$, we don't have $\alpha_{\rm II}$ and $\alpha_{\rm I\hspace{-.1em}I\hspace{-.1em}I}$. It is obvious that \cite{Ogasawara:2016yfk}
\begin{align}
\sin\alpha_{\rm II}=\sin\alpha_{\rm I\hspace{-.1em}I\hspace{-.1em}I}=1,\\ \,\cos\alpha_{\rm II}=\cos\alpha_{\rm I\hspace{-.1em}I\hspace{-.1em}I}=0.
\end{align}
So
\begin{equation}
\alpha_{\rm II}=\alpha_{\rm I\hspace{-.1em}I\hspace{-.1em}I}=\frac{\pi}{2}.
\end{equation}
By specific calculation, we can know $\alpha_{\rm IV}<\alpha_{\rm I\hspace{-.1em}I\hspace{-.1em}I}=\alpha_{\rm II}<\alpha_{\rm I}$, the escape probability of the spinning particle can thus be defined by
\begin{equation}
\rho\equiv\frac{\alpha_{\rm I}-\alpha_{\rm IV}}{2\pi}.
\end{equation}

\begin{figure*}[!htbp] 
   \centering
    \includegraphics[width=2.2in]{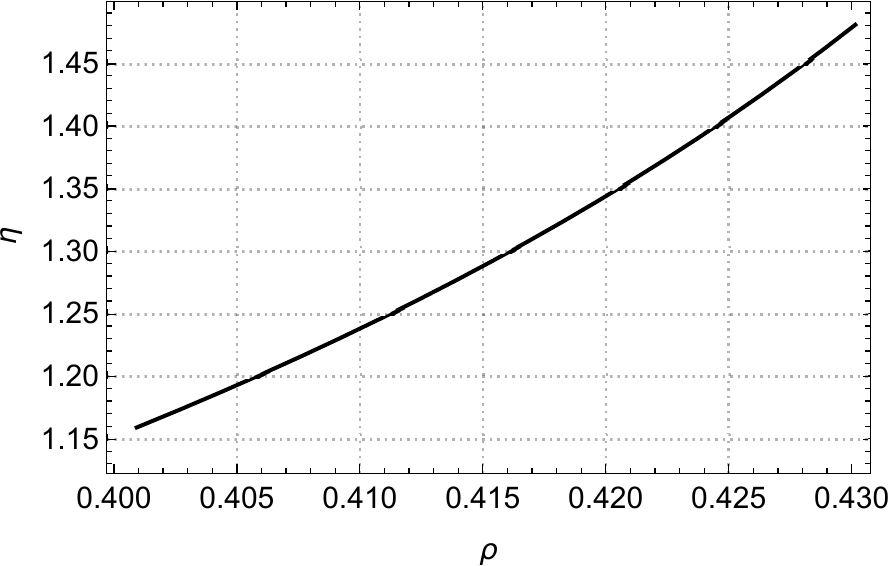}
        \includegraphics[width=2.2in]{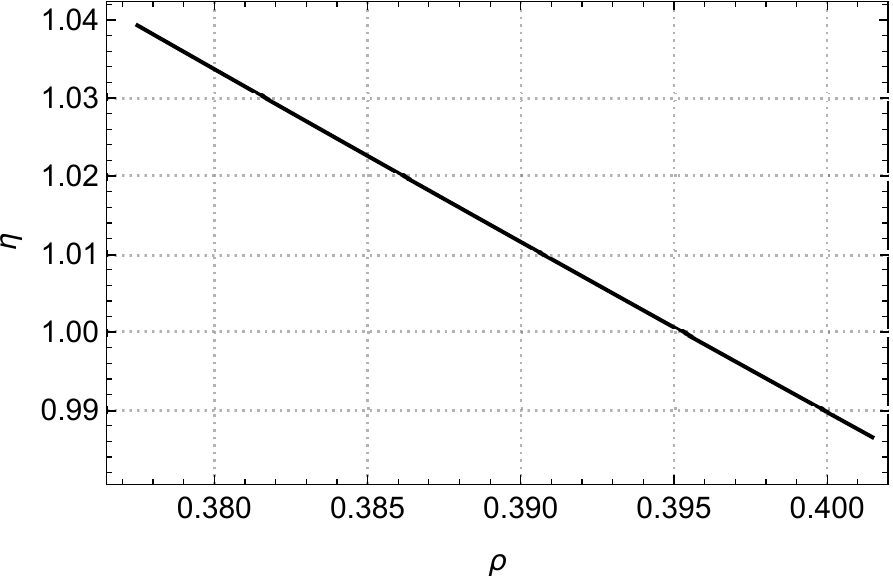}
         \includegraphics[width=2.2in]{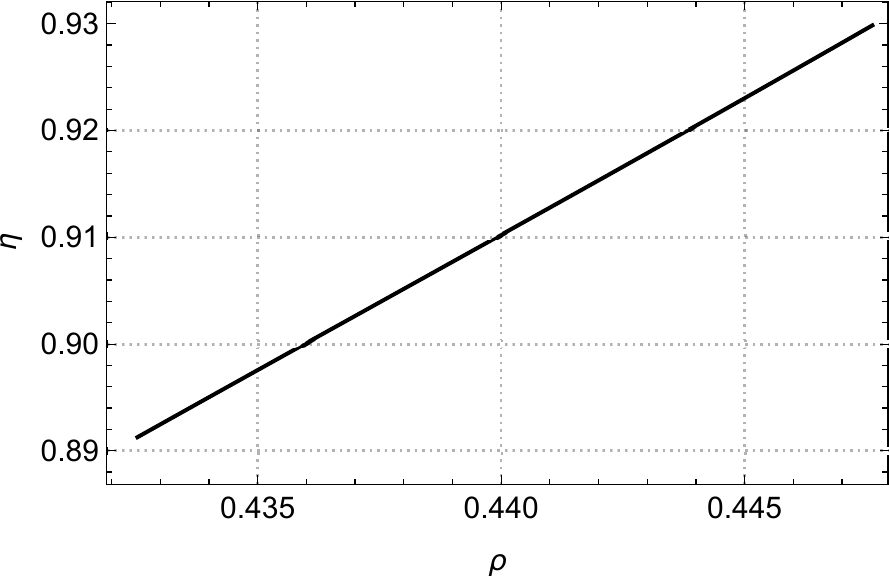}
   \caption{Variations of the energy extraction efficiencies in the collisional Penrose process with respect to the escaping probability of the produced particle for $M=1,\,\epsilon=0,\, a=1\,(\mathrm{left}),\,0.65\,(\mathrm{middle}),\,0.9\,(\mathrm{right}),\, r_{*}=1.01 \,r_{+} \,(\mathrm{left}),\,1.1\,r_+(\mathrm{middle}),\,1.35\, r_{+}\, (\mathrm{right}), j_{1}=j_{2}=2,\, e_{1}=e_{2}=1,\,-0.1<s_{0}<0.1$. Different from the cases in Fig. \ref{ep}, here the energy of the escaping particle cannot be set to be 1 and it depends on the collisional process.}
   \label{case}
\end{figure*}

For the extreme Kerr black hole, $j_{+}(r_+)=2 e$, we can analytically calculate the escape probability in the linear order of the particle's spin as
\begin{equation}\label{x1}
\rho=\mathcal{Z}_{1}+\mathcal{Z}_{2}s+\mathcal{O}(s^2),
\end{equation}
where
\begin{equation}
\mathcal{Z}_{1}=\frac{1}{2}-\frac{1}{2\pi}\arcsin\left(\frac{e}{\sqrt{e^2 (r+1)^2-r^2}}\right),\nonumber
\end{equation}
\begin{equation}
\mathcal{Z}_{2}=\frac{e^3 \left(r^3+2 r^2+2 r+1\right)-e r^3}{r \left[e^2 (r+1)^2-r^2\right] \sqrt{r \left[e^2 (r+2)-r\right]}}>0.\nonumber
\end{equation}
So we can know that the escape probability of the particle increases with the spin of the particle. This can be further confirmed in the left diagram of Fig. \ref{ep}, where we have taken the time-like condition and the  forward-in-time condition into consideration.
If $e=1$, we have 
$$\lim_{r_*\to 1} \rho\to \frac{1}{2}-\frac{\arcsin \left(\frac{1}{\sqrt{3}}\right)}{2 \pi }\sim 0.402$$
for the spinless particle, which is close to but not equal to the one $(\rho\sim 0.412)$ obtained in Ref. \cite{Ogasawara:2016yfk}. It is due to that we have chosen the Carter frame for the particle here yet locally non-rotating frame was used there. It tells us that the escape probabilities of the particles vary with the reference frame we choose. However, we reckon that different selections of observer's frames will not qualitatively change the results we report in this paper. Detailed investigations on this will be presented in \cite{Zhang:2020pay}.  At the same time, we can see that the escape probability of the particle is dependent upon its energy.

For the non-extreme Kerr black hole, we can numerically calculate the escape probabilities of the spinning particles. We individually show the variations of the escape probabilities with respect to the spin of the particles for cases of $r_{*}<r_{1}$ and $r_{*}>r_{1}$ in the middle and right diagrams of Fig. \ref{ep}. We can see that the escape probability increases with the spin for $r_{*}>r_{1}$ but decreases with the spin for $r_{*}<r_{1}$.

\section{The escape probability and the collisional Penrose process}\label{epcpp}
In this section, we will study the relation between the escape probability and the maximum energy extraction efficiency in the collisional Penrose process for the spinning particles. We consider that an outgoing particle 1 collides with an ingoing  particle 2 in the ergosphere of the Kerr black hole and suppose that the mass, the angular momentum $j_{k}$ and the energy $e_{k}$ of the two particles are equal when they collide.

The maximum energy extraction efficiency of this kind of process around the extreme Kerr black hole has been explored in Ref. \cite{Liu:2018myg} in case that the produced two particles are both massive. We at here will generally calculate the maximum energy extraction efficiency in both the non-extreme and extreme Kerr black hole backgrounds.

Not loss of generality, we denote the spin of both particle 1 and 2 as $s_{0}$, and suppose that both the produced outgoing massive particle 3 and the ingoing massive particle 4 are endowed with the same spin $s_{0}$ based on the conservation of the spin. The total radial momentum is conserved, which gives
\begin{equation}\label{condisix}
u^{(1)}_{1}-u^{(1)}_{2}=u^{(1)}_{3}-u^{(1)}_{4}=\epsilon,
\end{equation}
where $\epsilon$ denotes the total radial momentum of the particles. Due to the conservations of the energy and the angular momentum, we have
\begin{equation}\label{condifour}
e_1 +e_2 =e_3 +e_4=2e_{1},
\end{equation}
\begin{equation}\label{condifive}
j_1 +j_2 =j_3 +j_4=2j_{1}.
\end{equation}
Substituting $e_{4}=2e_{1}-e_{3}$ and $j_{4}=2j_{1}-j_{3}$ into (\ref{condisix}), we can obtain
\begin{equation}\label{afsuej}
u^{(1)}_{3}(a, e_{3}, j_{3},s=s_{0},r)-u^{(1)}_{4}(a, e_{1}, j_{1}, e_{3},j_{3},s=s_0,r)=\epsilon.
\end{equation}
Then we obtain
\begin{equation}\label{solalj}
j_{3}=j_{3}(a, e_{1},j_{1},e_{3}, s_{0}, r, \epsilon).
\end{equation}
Substituting it into the effective potential $O$ for particle 3 and using the conditions
\begin{align}
O_{3}(a, e_{1}, j_{1}, e_{3},s_{0},r, \epsilon)&\geqslant 0,\quad \\
\left.v_{(a)}v^{(a)}\right|_{a, s_{0}, r, \epsilon,e=e_{3},j=j_{3}}&<0,\\
\left.\frac{dt}{d\tau}\right|_{a, s_{0}, r, \epsilon,e=e_{3},j=j_{3}}&>0,
\end{align}
we have
\begin{align}
O_{3}&=O_{3}(a, e_{1}, j_{1}, e_{3},s_{0},r, \epsilon)=\mathcal{A}e_{3}^{2}+\mathcal{B}e_{3}+\mathcal{C}\geqslant 0,
\end{align}
where $\mathcal{A}<0$.
The physically reasonable maximum value of the escaping massive particle 3 is
\begin{align}
e_{3}=e_{3}^{\mathrm{max}}(a, e_{1},j_{1},s_{0},r,\epsilon)=\frac{-\mathcal{B}+\sqrt{\mathcal{B}^{2}-4\mathcal{A}\mathcal{C}}}{2\mathcal{A}},
\end{align}
and the maximum energy extraction efficiency $\eta$ of the collisional Penrose process is
\begin{equation}
\eta=\frac{e_{3}}{e_{1}+e_{2}}.
\end{equation}
For the produced escaping particle 3, we can also calculate its escape probability, which is related to its energy $e_3$. Then it is intriguing to analyze the relation between the escape probability of the particle and the efficiency of the energy extraction process.

We have shown the relation between the escape probability of the produced emitted particle and the maximum energy extraction efficiency of the collisional Penrose process in Fig. \ref{case}. From the diagrams, we can know that for the extreme Kerr black hole, the energy extraction efficiency increases with the increasing escape probability. For the non-extreme Kerr black hole, there are two different variation trends. If the radius of the collision point is less than the one which makes $j_+$ minimal, the energy extraction efficiency decreases with the increasing escape probability; if the radius of the collision point is greater than the one which makes $j_+$ minimal but less than $r_e$, the energy extraction efficiency increases with the increasing escape probability.

\begin{figure}[!htbp] 
   \centering
    \includegraphics[width=2.8in]{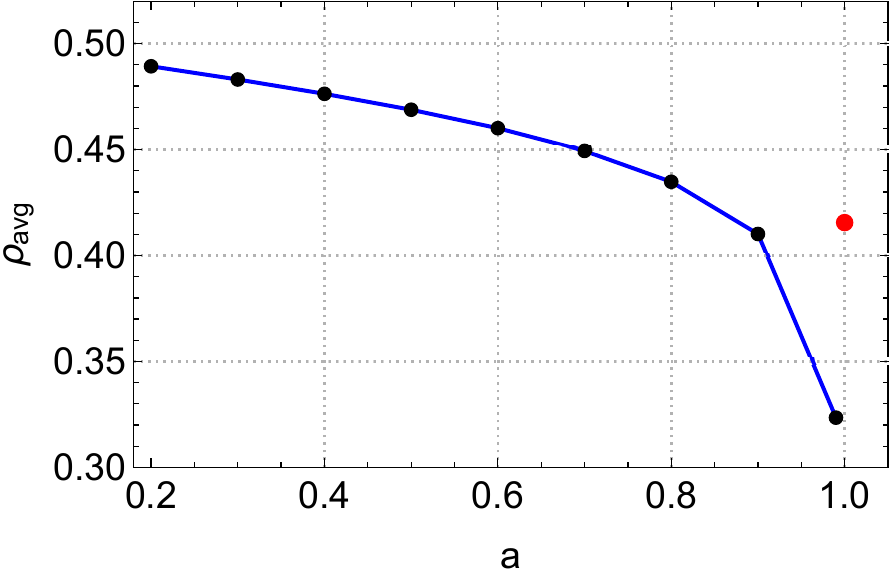}
   \caption{Variation of the average escape probability $\rho_{\mathrm{avg}}$ of the particle produced in the collisional Penrose process with respect to the rotation parameter  $a$ of the black hole  for $M=1,\,\epsilon=0,\,  j_{1}=j_{2}=2,\, e_{1}=e_{2}=1\,,-0.1<s_{0}<0.1$. The escape probability is roughly calculated by $[\eta(s_0=-0.1)+\eta(s_0=0.1)]/2$ as we have chosen $-0.1<s_0<0.1$ and $\eta$ almost linearly changes with $\rho$ and $s$.  As we have chosen $r_*=1.01\,r_+$, the rotation parameter of the black hole should be $a>0.198$ so that the collisional point is inside the ergosphere. Note that the red point corresponds to the extreme black hole case.}
   \label{case2}
\end{figure}

However, we can see that the particle's escape probability in the collisional Penrose process does not change significantly with the particle's spin. In this regard, different energy extraction efficiencies of the collisional Penrose process can almost correspond to an average escape probability of the spinning particle. Following this, we show the average escape probability $\rho_{\mathrm{avg}}$ of the particle produced in the collisional events which take place nearby the event horizon of the black hole in terms of the black hole rotation parameter in Fig. \ref{case2}. We see that the average escape probability of the spinning particle produced in the collisional Penrose process nearby the event horizon of the black hole decreases with the rotation parameter of the Kerr black hole, except for the extreme case (corresponding to the red point). There are subtle properties one should notice here. Because $r_*=1.01r_+$, we can see a jump of $\rho_{\mathrm{avg}}$ from $\rho_{\mathrm{avg}}(a=0.99)$ to $\rho_{\mathrm{avg}}(a=1)$. In fact, there is a turning point for the curve between $a=0.99$ and $a=1$, after which $\rho_{\mathrm{avg}}$ increases with $a$, as we will have $r_*>r_1$ if $a$ increases to a certain value very close to 1. For instance, we have $\rho_{\mathrm{avg}}(a=0.9999)=0.344$. Anyway, if we choose $r_*\to r_+$, we can obtain a monotonically decreasing curve from $a\to 0^+$ to $a=1$, as we always have $r_*\leqslant r_1$ and the ``$=$'' is for $a=1$.

\section{Conclusions}\label{con}

In this paper, we revisited the collisional Penrose process in term of the escape probability for the spinning particle. To this end, we first studied the law of the escape probability for the spinning particle around the Kerr black hole. We found that the escape probability $\rho$ of the spinning particle increases with the particle's spin $s$ around the extreme Kerr black hole. In the non-extreme Kerr black hole background, $\rho$ decreases with $s$ if the particle source locates at $r_*<r_1$ and $\rho$ increases with $s$ if the particle source locates at $r_*>r_1$, where $r_1$ is the position which makes the impact parameter of the particle minimal.

We then investigated the relation between the energy extraction efficiency $\eta$ of the collisional Penrose process and the escape probability $\rho$ of the produced particle with varying spin. Note that the escape probability of the particle is affected by the particle's energy, so we cannot obtain the law directly. By calculation, we discovered that $\eta$ increases with $\rho$ for the extreme Kerr black hole. However, for the non-extreme Kerr black hole,  $\eta$ decreases with $\rho$ if the collisional point locates at $r_*<r_1$ and $\eta$ increases with $\rho$ if the collisional point locates at $r_*>r_1$. 

Noticing that the change of the particle's escape probability is relatively minuscule, we further studied the average escape probability for the spinning particle produced in the collisional Penrose process. As a result, we found that the particle's escape probability decreases with the rotation parameter of the Kerr black hole in the horizon limit. 

Our discussion is based on viewing the particle as an extended object which has small varying spin. We can know that $r_1$ is a critical position where properties of the escape probability and the energy extraction efficiency change qualitatively for the non-extreme Kerr black hole. In the extreme Kerr black hole case, $r_1$ coincides with the event horizon. Our results will be beneficial to the astrophysical observation investigation. For the astrophysical relevant black holes, $a\lesssim 0.998$. Our results predict a near-horizon physical scenario around the astrophysical rotating black hole: (1) The escape probability of the spinning particle decreases with the pole/de-pole spin angular momentum of the particle; (2) the energy extraction efficiency decreases with the minuscully increasing escape probability of the spinning particle.

\section*{Acknowledgements}
Jie Jiang  is supported by the National Natural Science Foundation of China (Grants No. 11775022 and
11873044). Ming Zhang is supported by the Initial Research Foundation of Jiangxi Normal University with Grant No. 12020023.

\end{document}